\documentstyle[12pt]{article}
\begin{document}
\newcommand{\ket}[1] {\mbox{$ \vert #1 > $}}
\newcommand{\bra}[1] {\mbox{$ <
#1 \vert $}}
\newcommand{\bk}[1] {\mbox{$ \langle #1 \rangle $}}  \newcommand{\scal}[2]
{\mbox{$ < #1 \vert #2 > $}}
\newcommand{\expect}[3] {\mbox{$ \bra{#1} #2
\ket{#3} $}}  \newcommand{\ki}{\mbox{$ \ket{\psi_i} $}}
\newcommand{\bi}{\mbox{$ \bra{\psi_i} $}}  \newcommand{\p} \prime
\newcommand{\e} \epsilon 
\newcommand{\la}{\omega}

\newcommand{\om} \omega  \newcommand{\cc}{\mbox{$\cal C $}}
\newcommand{\al}{\mbox{$ \alpha $}}

\newcommand{\be}{\mbox{$ \beta $}}

\begin{flushright}
LPTENS 96/17 \\
TAU 2325-96\\
March 1996\\
\end{flushright}
\vskip .8cm
\vskip .8cm
\centerline{\bf  Comment on ``Vanishing Hawking Radiation from a Uniformly}
\centerline{\bf Accelerated Black
Hole'', P. Yi, Phys. Rev. Lett. 75 (1995) 382.}
\vskip 1.5cm
\centerline{S. Massar\footnote{e-mail: massar @ ccsg.tau.ac.il},
}
\centerline{ Raymond and Beverly Sackler Faculty of Exact Sciences,}
\centerline{ School of Physics and Astronomy,}
\centerline{Tel-Aviv University, Tel-Aviv 69978, Israel}
\vskip 5 truemm
\centerline{R. Parentani\footnote{Unit\'e propre de recherche du C.N.R.S.
associee \` a  l'\' Ecole
Normale Sup\'erieure et \` a l'Universit\' e de Paris Sud.
e-mail: parenta@physique.ens.fr}}
\centerline{Laboratoire de Physique Th\'eorique de l'\' Ecole
Normale Sup\'erieure,}
\centerline{24 rue Lhomond,
75.231 Paris CEDEX 05, France}
\vskip 3 truecm
\vskip 1.5 truecm

{\bf Abstract }

\noindent

We argue that the analysis of transients, of decoherence effects, or of any 
breaking of the exact boost invariance of the Ernst metric
shows that uniformly accelerated black holes do emit an
energy flux given by the Doppler-shifted Hawking radiation, in 
perfect {\it agreement} to what happens for accelerated 
particles.
\newpage

In the above letter, P. Yi used the structure of modes in the Ernst metric
expressed in a {\it Rindler}-type of coordinates
 to argue that uniformly accelerated black 
 holes do not
 emit particles when their Hawking temperature equals their Unruh temperature.
 This problem is closely 
related to the 
problem of the energy emitted by a 
uniformly accelerated detector in thermal equilibrium with
the Fulling-Davies-Unruh heat bath\cite{Unr}, much more 
than to the problem of the accelerated charge
mentioned by Yi.
In the 
detector case, the analogue of Yi's result is Grove's 
analysis\cite{Grow} of the detailed balance 
between absorbed and emitted {\it Rindler} quanta which
suggests that no energy flux is emitted in the mean. However a global
analysis\cite{Unruh2}\cite{AuMu}\cite{GO} shows 
that a net positive {\it Minkowski} energy is emitted.
Furthermore,
it is equal to the 
sum of the Doppler shifted energies of each 
transition of the detector\cite{MaPa}
thereby confirming Unruh's original idea that each interaction 
with the heat bath results in the emission of one Minkowski quantum. 
For the same {\it kinematical} reasons, 
a global analysis of the accelerated 
black hole system 
will also show that black holes emit positive energy. 

In both cases,
the vanishing of the Rindler energy
is guaranteed by the stationarity of the equilibrium. 
In the accelerated detector case, it
is realized by an exact compensation between the transitions of the
thermalized detector and the Bogoljubov transformation between Rindler
and Minkowski modes, see \cite{Grow}. In the black hole case, the
thermal equilibrium of the detector is replaced by the thermal
Bogoljubov transformation at the event horizon (eq. (13) of Yi's
letter). The stationarity is realized by the exact compensation
between Bogoljubov transformations at the event and at the acceleration
horizon (i.e. the equality of eqs. (13) and (14)). However, as in the
detector case, this is a condition for {\it Rindler}
equilibrium at the quantum level and not a criterion for the 
absence of emission of Minkowski quanta.
Indeed, upon neglecting the recoils
of the detector, 
one must take into account 
the transients which occur when the detector is set into acceleration 
or coupled to the radiation field when one wants to compute the total Minkowski
energy emitted. 
For the total Rindler energy, on the contrary, one may safely 
neglect these transients.
The apparent
contradiction between these results
is resolved
when one recalls that the Rindler energy,
measured in the accelerated frame,  differs from the
Minkowski energy by an exponentially {\it diverging} Doppler shift. 
Thus the vanishing of the first does not imply
the vanishing of the latter.

In this respect, it should be pointed out that the {\it cancellation} of the 
Bogoljubov transformations advocated by Yi
relies on the fact that there are two accelerated black holes. 
If only one black hole is present, 
then the general proof given in the appendix B of ref. \cite{MaPa} 
applies and the production of Minkowski quanta is inevitable.
(For the reader's convenience we have added this Appendix below).
To understand the 
misleading 
role of the second black hole, it is instructive to 
consider the scattering in $1+1$ dimensions by two accelerated mirrors 
on perfectly opposite trajectories $z^2-t^2 = a^{-2}$. By ignoring the 
particle content of the
singularity at $V=0$\cite{rec}, one may be tempted
to conclude that no Minkowski quanta are emitted in this case as well.

For the same reason, any breaking of the exact boost invariance of the
Ernst metric is also sufficient to invalidate Yi's cancellation.
For instance imagine that the trajectory of the left
black hole is shifted by $\Delta V_2 = \epsilon$. The acceleration horizons
for the left and right black holes will no longer coincide, and the modes near
the acceleration horizon take the form $\psi_L^{(\omega)}\simeq
\vert\kappa V_2 - \kappa\epsilon \vert^{i\omega/\kappa}$ and 
$\psi_R^{(\omega)}\simeq 
\vert\kappa V_2  \vert^{i\omega/\kappa}$.
This will lead to the emission of radiation at late times. Indeed upon
building wave packets ($
= \int d\omega e^{-i\omega v_0} 
f_{\omega_0}(\omega)\psi_{}^{(\omega)}$ 
where $f_{\omega_0}$ is a function of width $\Delta \omega$ centered 
around $\omega_0$),
one sees immediately that these wave packets differ from those 
built from the unperturbed 
modes
for $ v_0> \kappa^{-1}
 ln(\Delta \omega \epsilon)$, see eqs. (9,10).
Thus after this logarithmically short time, the Bogoljubov 
transformation is no longer given by eq. (14) and a 
steady flux of particles is 
emitted. 

Returning to the accelerated detector problem, 
we recall that when its position is quantized, the recoils induced 
by the interactions with the radiation
entail a decoherence
of the detector-field system\cite{rec} which simplifies the 
problem. Indeed, the decoherence implies that the
interferences (which ensured  that the energy flux is located only 
in the transients at $V=0$) 
are now inoperative, and a steady, positive and 
incoherent flux of energy is emitted.
This decoherence is an inevitable consequence of momentum conservation.
Indeed, 
upon absorbing or emitting a quantum of Rindler energy $\la$, 
the detector momentum changes
by $\delta P = \la e^{a \tau_0}$.  $\tau_0$ is the proper time at which
 the emission occurs and $e^{a \tau_0}$
is the Doppler shift.
Since a localized detector's wave function
has finite width $\Delta P$, after a logarithmically 
short time $a \tau_0 = ln(\Delta P / \la)$, these processes can
no longer interfere since the scattered wave function is orthogonal to the 
unscattered one.

Similarly, upon quantizing the position 
of the black hole,   
one shall also find that recoils induce decoherence 
and therefore a steady positive flux. The only assumption one
must make
is that quantum gravity be such that momentum conservation 
still holds for frequencies much smaller than Planckian ones. 
This assumption seems inevitable in the point particle limit (i.e.
when the radius of the hole goes to zero at fixed acceleration).
In that limit, the external 
geometry 
tends uniformly to the Melvin geometry 
and the scattering of the radiation field can no longer depend of the
internal geometry of the hole since the mean wavelength is
much large than its radius. 

In summary, the analysis of transients, of decoherence effects, or of any 
breaking of the exact boost invariance of the Ernst metric, 
show that black holes in uniform acceleration emit an
energy flux given by the Doppler-shifted Hawking radiation, in 
perfect {\it agreement} to what happens for point like particles.

\vskip 1.5 truecm

\newpage

\section{Appendix B of 
ref. [6]}

For any accelerated system coupled to the radiation
in such a way that the scattered radiation
modes are lineary related to the ingoing
modes, we prove that the elastic character
of the scattering process in the accelerated frame,
i.e. the absence of creation of Rindler quanta,
implies a production of Minkowski quanta.
In addition, we
believe
that the proof can be further generalized, 
using the same type of argumentation,
to nonlinear scattering processes.

The
proof goes as follow. Any linear scattering by an
accelerated
system in the right quadrant $(R)$ which does not lead to
the production of Rindler quanta
can be described
by
\begin{equation}
\tilde a_{\la,R} =  S_{\la \la^\prime} a_{\la^\prime,R}
 \label{vsix}
\end{equation}
where repeated indices are summed over.
 The matrix $S$ satisfy the
unitary relation
\begin{equation}
 S_{\la \la^{\prime\prime}}  S_{\la^{\prime\prime} \la^\prime}^\dagger
 = \delta _{\la \la^\prime}
 \label{vseven}
\end{equation}
which express the conservation of the number of Rindler quanta:
$S_{\la \la^\prime}$ mixes positive Rindler frequencies only.
It is convenient to introduce the matrix $T$ (from now on we
do not write the indices)
\begin{equation}
S=1+iT
 \label{veight}
\end{equation}
which satisfies
\begin{equation}
2 \mbox{Im} T = TT^\dagger
\label{vnine}
\end{equation}
We introduce also the vector operator $b= \left( a_{\la,R};a_{\la,L};
a_{\la,R}^\dagger;a_{\la,L}^\dagger \right) $ which contains the 
Rindler modes leaving in the left quadrant $(L)$.
 Then eq. (\ref{vsix}) can be written as
\begin{equation}
\tilde b= {\cal  S} b
\label{trente}
\end{equation}
where $ {\cal S}$ has the following block structure
\begin{equation}
 {\cal  S} =
\left( \begin{array}{cccc}
1+iT & 0 & 0 & 0 \\
0 & 1 & 0 & 0 \\
0 & 0 & 1-iT^\dagger & 0 \\
0 & 0 & 0 & 1
\end{array} \right)
\label{tone}
\end{equation}
since the 
modes on the left quadrant are still free. 

On the other hand, the Bogoljubov transformation 
which relates Minkowski and Rindler quanta reads in this
notation
\begin{equation}
c ={ \cal  B} b
\label{ttwo}
\end{equation}
where $c=  \left( a_{\la,M};a_{- \la,M};a_{\la,M}^\dagger;a_{- \la,M}^\dagger
\right)$ are the Minkowski operators, i.e. 
the destruction operators $ a_{\pm \la,M}$ 
annihilate Minkowski vacuum.
 $ {\cal  B}$ is
\begin{equation}
  {\cal  B} =
\left( \begin{array}{cccc}
 \alpha &  0 & 0 & - \beta \\
 0 & \alpha & - \beta  & 0 \\
 0 & - \beta &\alpha & 0 \\
 - \beta  & 0 & 0 & \alpha \\
\end{array} \right)
\label{tthree}
\end{equation}
The diagonal matrices (in $\la$)
$\alpha $ and $\beta$ have been taken real.

Then, from eq. (\ref{trente}) and eq. (\ref{ttwo}),
the scattered Minkowski operators $\tilde c$ are given in terms of the
ingoing operators $c$ by the following matrix  relation
\begin{equation}
\tilde c =  {\cal  B}{ \cal  S}{ \cal  B }^{-1} c = \left(
{ \cal  S} + {\cal  B}\left[ { \cal
S},{ \cal  B }^{-1} \right]_- \right) c = {\cal  S}_M c
\label{tfour}
\end{equation}
Since  ${\cal  S}$ and  ${\cal  B}$ do not commute,  ${\cal  S}_M$ has
non-diagonal elements: 
\begin{equation}
  {\cal  S}_M  =
\left( \begin{array}{cccc}
 \tilde \alpha_1 &  0 & 0 & - \tilde \beta _1 \\
 0 & \tilde  \alpha_2 & \tilde  \beta^\dagger _1  & 0 \\
 0 &  \beta^\dagger_2 &\alpha _1^\dagger & 0 \\
 - \tilde  \beta_2  & 0 & 0 &\tilde  \alpha^\dagger_2 \\
\end{array} \right)
\label{tfive}
\end{equation}
$\tilde \alpha$ $\tilde \beta$ are given in terms of $T$ by
\begin{eqnarray}
 \tilde  \alpha _1 &=&  1+i  \alpha T  \alpha
\nonumber\\
 \tilde
\beta _1  &=& -i  \alpha T \beta
\nonumber\\
\tilde  \alpha _2 &=& 1 +i \beta T^\dagger  \beta
\nonumber\\
\tilde
\beta_2 &=& i  \beta  T  \alpha
\label{tsix}
\end{eqnarray}
The non-diagonal matrix elements, $\tilde \beta$, 
mixe creation and destruction operators, and encode,
as usual, the amplitudes of pair creation.

Therefore, the
noncommutativity of ${\cal  S}$ and  ${\cal  B}$ is sufficient
to deduce that any scattering giving rise no production
of Rindler quanta
necessarily induces the pair production of Minkowski quanta.
If furthermore, the Rindler scattering is stationary during a lapse of proper
time much greater than $1/a$, that is, $S_{\la \la^\prime}
$ is, to a good approximation, diagonal in $\la$, then, the number of
created pairs of Minkowski quanta is proportional to the interval of
proper time, see \cite{MaPa} for the proof.


\begin{thebibliography}{999}



\bibitem{Unr}  W. G. Unruh, Phys. Rev. D {\bf  14} (1976) 870

\bibitem{Grow}P. Grove, Class. Quantum Grav. {\bf  3} (1986) 801

\bibitem{Unruh2}Unruh W. G., Phys. Rev. D {\bf 46} (1992) 3271

\bibitem{AuMu}Audretsch J. and M\"uller R., Phys. Rev. D {\bf  49} (1994)
 4056; Phys. Rev D {\bf 49} (1994) 6566;
Phys. Rev A {\bf 50} (1994) 1755

\bibitem{GO} R. Brout, S. Massar, R. Parentani and Ph. Spindel,
Phys. Rep.  {\bf 260}  (1995) 329.

\bibitem{MaPa}Massar S. and Parentani R., {\em From Vacuum Fluctuations to
Radiation: Accelerated Detectors and Black Holes}, 
gr-qc/ 9502024

\bibitem{rec}R. Parentani,
Nucl. Phys. B {\bf 454} (1995) 227 and hep-th/9509104.



\end{thebibliography}
\end{document}